\documentstyle[preprint,aps,version2]{revtex}
\begin{document}
\draft

\begin{title}
The influence of negative-energy states on proton-proton bremsstrahlung
\end{title}

\author{F. de Jong$^{1,2,3,}$\thanks{present adress: Inst. fuer Theoretische
Physik, Universitaet Giessen, 35362 Giessen, Germany}
, K. Nakayama$^{1,2}$}

\begin{instit}
$^{(1)}$Department of Physics and Astronomy, University of Georgia,
Athens, GA 30602\\
$^{(2)}$Institut f\"ur Kernphysik, Forschungszentrum J\"ulich,
52428 Germany\\
$^{(3)}$Kernfysisch Versneller Instituut, 9747 AA Groningen,
the Netherlands
\end{instit}

\begin{abstract}
We investigate the effect of negative-energy states on proton--proton
bremsstrahlung using a manifestly covariant amplitude based on a T-matrix 
constructed in a spectator model. We show that there is a large cancellation 
among the zeroth-order, single- and double-scattering diagrams involving 
negative-energy nucleonic currents. We thus conclude that it is essential to 
include all these diagrams when studying effects of negative-energy states. 
\end{abstract}

\section{Introduction}

Recent efforts in the study of the proton-proton bremsstrahlung ($pp\gamma$)
reaction have been directed towards investigating the various reaction 
mechanisms of this process at and beyond the pion production threshold energy
\cite{FdJ_ppg1,Jetter2,Eden2,FdJ_ppg2,FdJ_ppg3,Liou1,Eden3}. Among various
higher-order (in the photon momentum) processes studied so far, the most 
important contribution turns out to come from $\Delta$-decay diagrams 
(containing an $N \Delta \gamma$ vertex). Even at pion-threshold this 
contribution can enhance the cross-section up to 30\%
\cite{FdJ_ppg1,Jetter2,Eden2,FdJ_ppg2}. Another higher-order process stems 
from vector-meson-decay vertices. Of these, the $\omega$-decay into a pion 
and a photon is the most important one. These contributions have an effect 
opposite to that of $\Delta$-decay diagrams, i.e., they reduce the $pp\gamma$ 
cross section. However, their strength is only about one third of the 
 $\Delta$-decay diagrams \cite{FdJ_ppg3}. 
An important aspect of the $\Delta$- 
and meson-decay diagrams is that at pion-threshold their effect is due to 
interference with the dominant positive-energy nucleonic current (NN$\gamma$ 
vertex) contribution. The absolute value of these diagrams is small. The 
absolute value of the $\Delta$-decay diagrams only becomes comparable to the 
nucleonic contribution at higher energies. In addition to these higher-order 
processes, the question of pseudo-scalar/pseudo-vector(ps/pv) mixing of the 
 $\pi$NN coupling has been addressed in connection with the $pp\gamma$ 
reaction \cite{Liou1,Eden3}.

In this work, which is an effort \cite{FdJ_ppg1,FdJ_ppg2,FdJ_ppg3} to better 
understand the role of various reaction mechanisms contributing to the 
$pp\gamma$ process, we study the effects of nucleon negative-energy states. 
The investigation of negative energy states in $pp\gamma$ reactions has 
received limited attention \cite{Muendel,Eden3} because its contribution is 
also of higher-order in the photon momentum. 
In addition, since the most popular 
realistic NN interactions do not contain negative-energy states dynamically, 
it is difficult to add this contribution to $pp\gamma$ calculations based on 
these interactions in a consistent way, specially, satisfying charge 
conservation. We note that beyond the soft-photon approximation the negative 
energy contribution is essential for obtaining a fully gauge invariant 
$pp\gamma$ amplitude. An effort to achieve this consistency is reported in the
work by M\"undel\cite{Muendel}, who uses the Bonn potential to explore the 
role of nucleon negative-energy states. 
This contribution is taken into account 
through an effective two-body current in a $p/m$ expansion and considering the
$pp\gamma$ amplitude only in the zeroth-order scattering. M\"undel finds that 
this can interfere effectively with the positive-energy nucleonic 
contribution,
leading to about 10\% effects on the cross section. Eden and Gari\cite{Eden3} 
use the Ruhrpot potential in their $pp\gamma$ calculations. The authors 
\cite{Eden3} report a relatively large contribution from negative-energy 
states, which they can control by tuning the ps/pv ratio of the $\pi$NN 
coupling. However, although they do not use a $p/m$ expansion of the effective
two-body current, they also include this contribution only in the zeroth-order
scattering. In doing so, contributions from the single- and double-scattering 
(rescattering) diagrams are ignored. As we will point out later, when these 
diagrams are added to the zeroth-order scattering diagram, the total 
contribution to the cross section from nucleon negative-energy states 
practically vanishes due to a large cancellation among these diagrams.

Apart from the Ruhrpot T-matrix, we are aware of two NN T-matrices that 
include nucleon negative-energy states \cite{Fleischer,Gross}. In the present 
work we will use the T-matrix of Gross {\it et al}\cite{Gross}. First we will
discuss shortly this model for the NN interaction; secondly, we will discuss 
how we use this T-matrix in our model for the $pp\gamma$ reaction and, 
finally,
we will present results and compare them with other calculations and with the 
Triumf data of Ref.\cite{Michaelian}. It should be mentioned that, in contrast
to our previous work \cite{FdJ_ppg1,FdJ_ppg2,FdJ_ppg3}, we do not consider the
contribution of the $\Delta$-decay diagrams in this work, since there are no 
T-matrices available that include negative-energy states \`and $\Delta$ 
degrees of freedom. 
This is beyond the scope of the present work. 
Moreover, we believe
that the role of nucleon negative energy states in $pp\gamma$ reactions can be
explored ignoring the $\Delta$-decay diagrams, at least in this initial study.
The contribution of vector-meson-decay diagrams is also ignored in the present
work.

\section{Theoretical Framework}

The formalism we use to describe the hadronic system leading to our NN 
T-matrix is described in detail in Ref.\cite{Gross}. Starting from a 
field-theoretic formulation one employs a three-dimensional reduction of the 
Bethe-Salpeter equation that restricts one of the interacting particles to be
on its mass-shell (the other particle is allowed to be off-shell) and finds a 
manifestly covariant expression for the T-matrix which is symmetrized with 
respect to the on- and off-shell particles. The resulting equation is also 
known as the spectator equation. The propagator of the off-shell particle is 
retained in full, i.e., both the negative- and positive-energy parts of the 
propagator are kept. Gross {\it et al.} \cite{Gross} present four different 
models, all giving a good fit to the NN data below pion threshold with a 
$\chi^2$ that is comparable with those of the Paris and the Bonn potentials. 
In this work we use the IA model whose underlying interaction has only four 
mesons ($\pi, \rho, \sigma, \omega$) and a very limited number of free 
parameters. Apart from the coupling constants and the $\sigma$-mass, there are
two cut-off parameters and a parameter describing the ps/pv ratio of the 
$\pi$NN coupling. The specific T-matrix we use is calculated 
{\it independently} from Ref.\cite{Gross}. It is expressed in a plane-wave 
basis and reproduces the phase-shifts as reported in Ref.\cite{Gross}.

In the model we use for the hadronic interaction, phenomenological formfactors
are introduced on the meson-NN vertices. The use of phenomenological 
formfactors poses immediately a problem of introducing the electromagnetic 
interaction without violating current conservation. Gross and 
Riska \cite{Gross2} presented a method describing how this can be done; 
although it does not yield unique electromagnetic couplings to hadrons 
(this is impossible with phenomenological formfactors where the underlying 
structures are unknown), the method is quite general. 
We, therefore, follow their 
approach: interpret the formfactors as self-energies in the hadron propagators
and demand that the electromagnetic vertices obey the corresponding 
Ward-Takahashi identities. To achieve this, the hadronic formfactors are 
chosen to be separable, giving a meson-NN vertex
\begin{equation}
\Gamma_{{\rm NNmeson}} = h(p^2) h(p'^2) f(q^2) 
\Gamma^R_{{\rm NNmeson}},
\end{equation}
with $p$, $p'$ the momenta of the incoming and outgoing nucleon, respectively,
and $q=p'-p$, the momentum of the meson. $h(p^2)$ is the nucleonic formfactor 
(nucleonic in the sense that it depends only on the nucleon momentum $p$), 
$f(q^2)$, the mesonic formfactor and $\Gamma^R_{{\rm NNmeson}}$, the reduced 
meson-NN vertex. To retain the unit-residue at the poles of the nucleon and 
meson propagators one has the restriction on the formfactors, $h(p^2 = m^2_N) 
= 1$ and $f(p^2 = m^2_{\rm meson}) = 1$. In addition, both $h(p^2)$ and 
$f(q^2)$ should decrease at least like a power of their arguments as they 
approach infinity and have no zeros. In $pp\gamma$ reactions we do not have 
meson-exchange currents, and thus, the mesonic formfactor $f(q^2)$ does not 
enter in the discussion of gauge invariance. Recall that we do not consider 
vector-meson-decay contributions in this work. The nucleonic formfactor leads 
to the Ward-Takahashi identity for the $pp\gamma$ vertex
\begin{equation}
k_\mu \Gamma_R^\mu(p', p) = 
e \left( S^{-1}(p') - S^{-1}(p) \right) ,
\label{WTI}
\end{equation}
where $k = p - p'$ denotes the photon momentum, $\Gamma^\mu_R$ , the reduced 
$pp\gamma$ vertex and $e$, the proton charge. $S(p)$ is the modified nucleon 
propagator which includes the formfactor as a self-energy. It is expressed in 
terms of the Feynman propagator $S_F(p)=1/(\not \! p - m_N)$ as $S(p) = 
h^2(p^2) S_F(p)$.

Now, starting from the most general NN$\gamma$ vertex \cite{Bincer}, we can 
uniquely construct the longitudinal part of the $pp\gamma$ vertex for a real 
photon which satisfies Eq. (\ref{WTI}),
\begin{eqnarray}
& &\Gamma_R^\mu(p', p) = \nonumber \\
& &\frac{-ie} {{p'}^2 - p^2} 
\left[ \gamma^\mu
\left( \frac{p'^2}{h^2(p'^2)} - \frac{p^2}{h^2(p^2)}
\right) +
\left(\not \! p' \gamma^\mu \mbox{$\not \! p$}  +
m_N \mbox{$ \not \! p'$} \gamma^\mu + m_N \gamma^\mu \mbox{$\not \! p$}
\right)
\left( \frac{1}{h^2(p'^2)} -  \frac{1}{h^2(p^2)}
\right)
\right].
\label{mod_vertex}
\end{eqnarray}
This vertex reduces to the conventional one, $-ie \gamma^\mu$, when 
$h(p^2) = 1$. Also, note that the vertex is finite when $p' = p$. The
Ward-Takahashi identity does not provide a constraint on the transverse part 
of the vertex, and in particular, the effect of the meson-NN formfactors
on the magnetic part of the $pp\gamma$ vertex. To remove this ambiguity one 
needs to calculate the formfactors in a microscopic model, as is done e.g. in 
Ref.\cite{Doenges}. Therefore, we retained the conventional magnetic vertex,
\begin{equation}
\Gamma^\mu_{\rm mag}(p', p) = - e \kappa 
\frac{\sigma^{\mu \nu} k_\nu} {2 m_N},
\label{mag_vertex}
\end{equation}
with $\kappa=1.79$, the anomalous magnetic moment of the proton. Note in the
above equation that the electromagnetic formfactor as discussed, e.g., in 
Ref.\cite{Herrmann2} is also set to its on-shell value. Our full $pp\gamma$ 
vertex is, therefore, the sum of the vertices given by Eqs.
(\ref{mod_vertex}) and (\ref{mag_vertex}),
\begin{equation}
\Gamma^\mu = \Gamma_R^\mu + \Gamma^\mu_{\rm mag} \ ,
\label{full_vertex}
\end{equation}
and satisfies, by construction, the charge conservation requirement imposed by
the particular NN interaction used. 
 
The pp bremsstrahlung amplitude is obtained by sandwiching the vertex given by
Eq. (\ref{full_vertex}) with the two-nucleon wave functions calculated 
within the spectator model. 
These wave functions are expressed in terms of the T-matrix 
described in the beginning of this section. Note that, in this way, the 
electromagnetic interaction is taken into account to first-order in the 
coupling, while the strong interaction is taken to all orders. 
The resulting amplitude is shown diagramatically in Fig.\ref{diagrams1} and 
is the sum of the single-scattering diagrams, (a) and (b), and the rescattering
diagram, (c). 
The cross in Fig. (\ref{diagrams1}) indicates where the particle is restricted 
to be on its mass-shell; the $\pm$ denotes intermediate states which are 
off-shell and have both positive- and negative-energy components. 
The T-matrix is {\it symmetrized} with respect to the on- and off-shell 
particles, leading to an amplitude that fullfils the Pauli-principle. 
Note that our single-scattering diagrams, Figs. \ref{diagrams1}a,b, incorporate 
the zeroth-order scattering diagrams mentioned in the introduction 
(Figs. \ref{diagrams1}a,b with the T-matrix replaced by the potential). 
 
Now, in order to obtain the complete $pp\gamma$ amplitude within the spectator 
model, one has to add the rescattering diagrams Figs. \ref{diagrams1}d,e to 
diagrams Figs. \ref{diagrams1}a-c. 
This is peculiar to this type of model in which one of the interacting 
particles is restricted to its mass-shell. 
Since photons cannot couple to an on-shell particle, these diagrams are 
necessary in order to account for the complete $pp\gamma$ 
amplitude \cite{Gross2}. 
They are also needed to ensure gauge invariance. 
It should be stressed that the complete amplitude obtained in this way is 
manifestly covariant. 

Unfortunately, the diagrams Figs. \ref{diagrams1}d,e are very difficult to 
calculate. A close examination, however, reveals that the sum of them can be 
well approximated by the diagram Fig. \ref{diagrams1}c. We, 
therefore, calculate the diagram Fig. \ref{diagrams1}c and multiply it by a 
factor of 2 in order to account for the total rescattering diagrams in the 
present work, i.e., (c)+(e)+(d) $\approx$ 2$\times$(c). 
Of course, this approximation destroys 'exact' gauge invariance of the 
complete amplitude. 
However, the violation is minor for we find the contraction of our 
approximated complete 
amplitude with the photon momentum, $k_\mu M^\mu$, reasonably close to zero. 
This means that we have a reasonable numerical fulfillment of gauge invariance.
A positive point in the use of a spectator-type three-dimensional reduction is
that it allows us to treat the energy variable in all diagrams consistently. 
The point is that it is kinematically impossible to treat the interacting two 
nucleons symmetrically when a photon couples to one of them. Therefore, it is 
difficult to use consistently in NN bremsstrahlung calculations those 
T-matrices obtained using symmetric three-dimensional reductions, such as those
based on the Blankenbecler-Sugar reduction. We feel that the correct treatment
of this aspect is a nice feature of the spectator model. 

\section{Results}
 
With the model described above we performed calculations for a selected set
of kinematics from the coplanar TRIUMF experiment \cite{Michaelian}. The 
results are shown in Fig. \ref{results_1}. The dotted lines denote the full 
results, including all diagrams with both positive- and negative-energy state 
contributions; the long-dashed lines are calculated by restricting the 
 $pp\gamma$ vertex to positive-energy states only (positive-energy nucleonic 
current). Of course, in this case, negative-energy states are still present in
the intermediate states of the T-matrix. For comparison, we also show the 
results based on the Bonn OBEPQ potential\cite{Machleidt} with only the 
positive-energy nucleonic contributions (solid lines). Comparing these results
(long-dashed and solid) we observe that they are similar to each other and 
fall into the trend observed in earlier works, i.e., T-matrices that fit the 
phase-shifts give similar results for the $pp\gamma$ observables. We now find 
the same even when the T-matrix includes negative-energy particles in its 
intermediate states. Also, recall that due to different meson-NN formfactors
we have to use different $pp\gamma$ vertices for the spectator model and Bonn 
T-matrices. Our results indicate that these formfactors have little influence
on the observables when the photon couples to a positive-energy nucleon,
although this is not the case when the photon couples to a negative energy 
nucleon as will be discussed later. The most remarkable finding, however, 
appears when we compare the results with and without the negative-energy 
nucleonic current (dotted and long-dashed curves, respectively) for the cross 
section. As can be seen, the negative-energy nucleonic current shows 
practically no influence on this observable. This is not due to the individual
diagrams in Fig. \ref{diagrams1} being small; on the contrary, they are large. 
But the sum of all the negative-energy diagrams has no effect on the cross 
section.

To detail the last point further, we show in Fig. \ref{results_2} the results
when only the single-scattering diagrams are considered with (dashed lines)
and without (solid lines) the negative-energy nucleonic current. As can be 
seen, the effect of the negative-energy state on the cross-section from the
single-scattering diagrams is large. A closer examination of the matrix 
elements shows that this large contribution stems from very large $M^{\mu=1}$ 
and $M^{\mu=2}$ matrix elements of the $pp\gamma$ amplitude in the 
 $++ \rightarrow +-$ spin transition channel. This contribution stems from the 
part of the $pp\gamma$ vertex given by Eq. (\ref{mod_vertex}). However, in this 
channel the sum of all the negative-energy single-scattering contributions is 
almost exactly cancelled by the sum of the negative-energy rescattering 
diagrams, leading to the observed nearly null-result for the cross-section in 
Fig. \ref{results_1}. Unfortunately, up to now we have not been able to find a 
specific explanation as to why this cancellation should occur, nor why these 
particular spin transition matrix elements should be large. Gauge invariance 
appears to be important here, for the cancellation is much less perfect when
we use the conventional vertex ($-ie\gamma^\mu$) in conjunction with the Gross 
T-matrix instead of the modified one given by Eq. (\ref{mod_vertex}) as 
required by gauge-invariance. 
Note that we also constructed a Bonn T-matrix with an 
outgoing negative energy state. Using this T-matrix we observed exactly the 
same feature as mentioned above. This indicates that the observed cancellation
may be a general feature of calculations based on potential models.  

In contrast to the cross section, the analyzing power is much more affected by
the negative-energy contribution as is shown in Fig. \ref{results_1}. Already 
the result with only the positive-energy nucleonic current (long-dashed line) 
shows some deviation from the corresponding Bonn result (solid line). This 
deviation takes the result away from the data. However, this is not too 
discomforting since other effects, like including the $\Delta$-isobar, tend to
push the analyzing power towards the data 
\cite{FdJ_ppg1,Jetter2,Eden2,FdJ_ppg2}. Including the negative-energy nucleonic
current (dotted curve) the minimum in the analyzing power at intermediate 
photon angles fills up almost completely. With this result it is difficult to 
imagine an effect that would push the curve back onto the data. The effect of 
the negative-energy nucleonic current on the analyzing power when only the 
single-scattering diagrams are considered is also illustrated in 
Fig. \ref{results_2}. Again, its effect is large and comparing with the result 
in Fig. \ref{results_1} we see that there is a large cancellation among the
diagrams involving negative energy states. In view of this, it is possible 
that a more refined calculation will give better results for the analyzing 
power, since in the present calculation we 'lost' exact gauge invariance due 
to the approximation made for evaluating the $pp\gamma$ amplitude, and that, 
we might miss higher-order cancellations among the negative-energy 
contributions. This has to be investigated in the future. 

At this point, we note that the above finding forces us to revise the claim 
made in Refs. \cite{Liou1,Eden3} that one can determine the ps/pv content of 
the $\pi$NN vertex by means of the negative-energy contributions to $pp\gamma$ 
observables. They take into account the negative-energy nucleonic current 
contributions only in the zeroth-order scattering and, thus, miss the 
cancellation observed in the present calculation. Of course, this does not 
necessarily imply that the $pp\gamma$ reaction is not sensitive to the ps/pv 
mixing. It simply means that one should be careful in drawing conclusions 
based on calculations that include only the zeroth-order scattering diagrams.

\section{Summary and Conclusions}

We investigated the role of nucleon negative-energy states on the $pp \gamma$ 
process. We used a model in which the negative energy states are treated 
dynamically in the hadronic interaction and the electromagnetic interaction is
introduced consistently with the hadronic interaction so that the current
conservation is satisfied. The photon is allowed to couple to both positive
and negative energy nucleons. For cross sections we found almost no effect 
of the negative energy states. We showed that this is not due to the separate 
diagrams being small, but caused by an almost perfect cancellation among all 
diagrams involving negative-energy nucleonic currents. Therefore, we concluded
that it is crucial to include all negative-energy diagrams in the calculation.
The results for the analyzing power are less encouraging when compared to the 
TRIUMF data \cite{Michaelian}. However, we believe that there are possible 
refinements in the model which might improve upon this. If the effect of the 
negative energy states on the analyzing power found in this work is genuine, 
i.e., not caused by our approximate treatment of gauge invariance due to 
numerical difficulties, reproducing the analyzing power as measured in the 
TRIUMF data will be a severely restrictive test on potential models of 
pp-bremsstrahlung reactions which incorporate nucleon negative energy states. 
Certainly, there is much to be done before this reaction mechanism is throughly
understood.

Finally, we mention that investigation of the ps/pv ratio of the $\pi$NN 
coupling using the $pp\gamma$ reaction as suggested in Refs.\cite{Liou1,Eden3}
should be revised. The sensitivity of the $pp\gamma$ reaction to this ratio 
arises basically from the negative-energy state contribution. Therefore, as we
have shown in this work, one has to be cautious with the conclusions based on 
calculations where the negative-energy nucleonic current is accounted for only
in the zeroth-order scattering \cite{Liou1,Eden3}.

\acknowledgments

We thank prof. Gross for valuable comments and Gary Love for a careful 
reading of the manuscript. This work was 
supported in part by COSY, KFA-J\"ulich, grant nr. 41256714.

\figure{
\label{diagrams1}
Diagramatic representation of the $pp\gamma$ amplitude in the present model.
(a) and (b) are the single-scattering diagrams; (c)-(e), the rescattering 
diagrams. The cross indicates an on-shell intermediate state, the $\pm$ is an 
off-shell intermediate state which has both positive and negative energy 
contributions. Diagrams (d) and (e) are peculiar to the present model.
}

\figure{
\label{results_1}
Results for the cross-section and analyzing power. The dotted line stands
for the full calculation including all diagrams; the dashed line with only
positive-energy diagrams. The solid line represents the Bonn OBEPQ results.
The data are from Ref. \cite{Michaelian}; the cross-sections were {\it not} 
multiplied by a factor 2/3; the analyzing power is multiplied by -1.
}

\figure{
\label{results_2}
Results for the cross-section and analyzing power. The solid line is
the result with only positive-energy diagrams. The dashed line is the
result when only the single-scattering negative energy diagrams are
added. The dotted line is the full result which, for the cross-section, 
coincides with the solid line. 
}

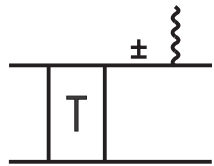
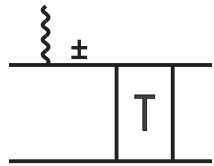
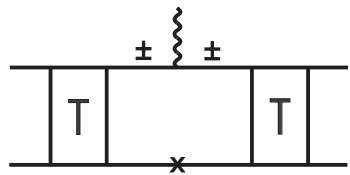
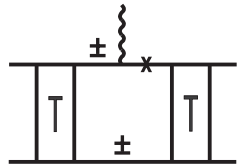
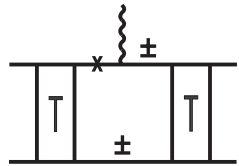
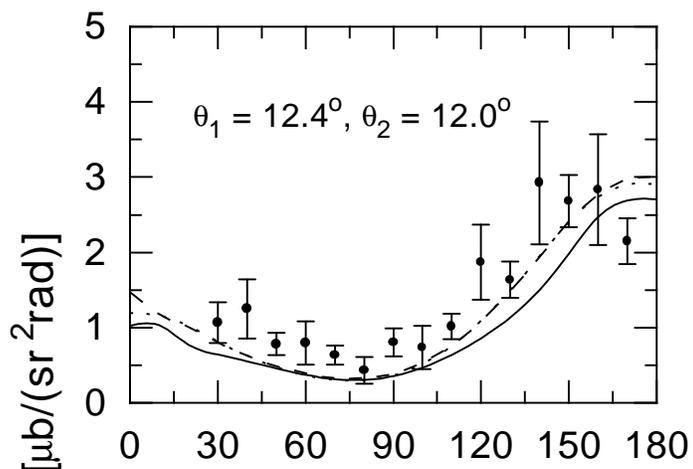
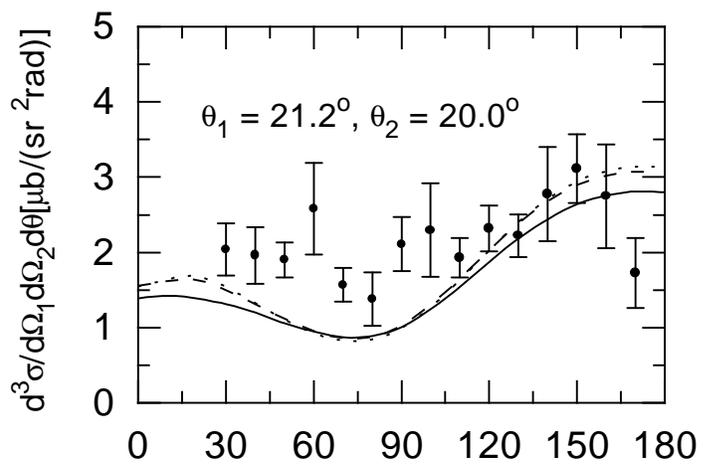
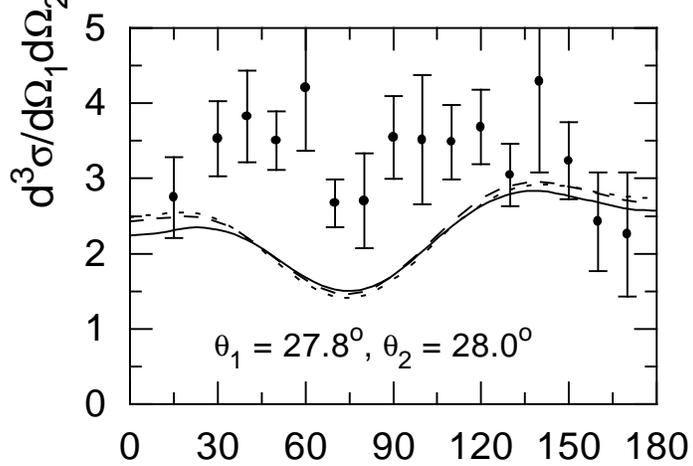
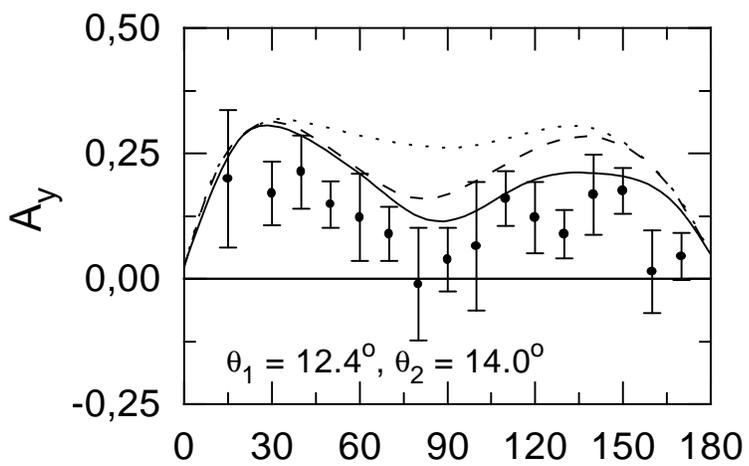
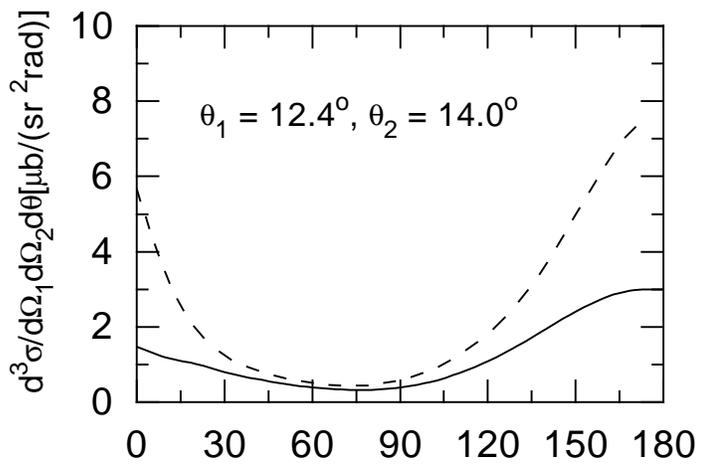
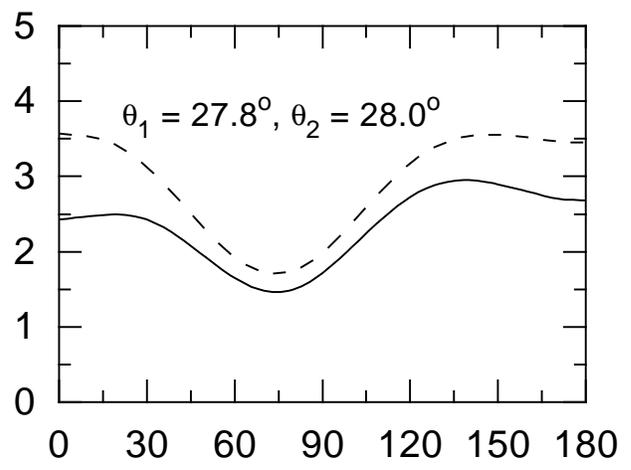
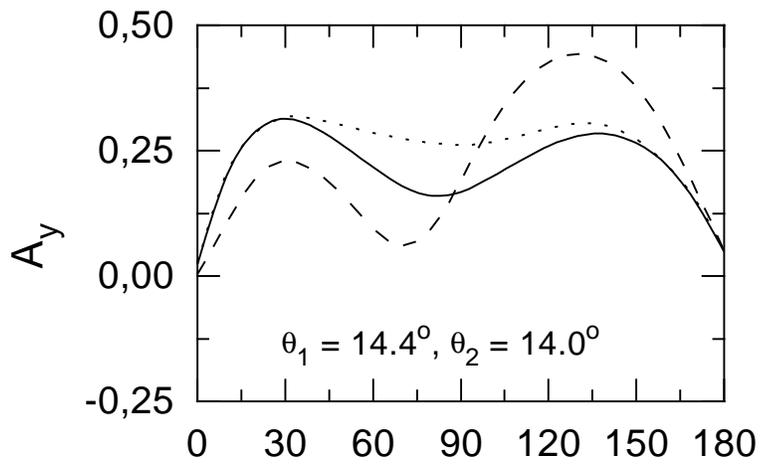
\end{document}